# The Use of NMR for Solving the Problems of Quantum Computers

Voronov V.K. Irkutsk State Technical University, Irkutsk, Russia

Though the idea of the possibility of making quantum computers was expressed by R. Feynman more than 15 years ago the actual results of its implementation have been obtained during the recent two-three years (see [1,2,3] and the literature given here). The basic stage of making quantum computers is searching for such physical processes which are made at an atomic and molecular level and which would eventually ensure the realization of logic operations similar to those in classical computers. Consequently, the progress in making quantum computers is largely defined by the level of our knowledge about the structure and dynamically multielectronic quantum (atomic and molecular) systems. Nowadays a nuclear magnetic resonance (NMR) is the most perspective in this respect. It is with the help of NMR that quantum algorithm was implemented for the first time [1]. For the sake of justice it should be noted that the modern state of the theory and NMR experimental security makes it possible to consider it as one of the best methods for obtaining information on the structure and behavior of quantum systems. It follows that the solution of, at least, some problems facing the creation of quantum computers. In particular, the Toffoley's gate-an important element of the calculation procedure can be simulated in NMR by the system of spins ABC characterized by the parameters: $\delta_A, \delta_B, \delta_C$; $J_{AB} \neq 0$; $J_{AC}(J_{CB}) \neq 0$; $J_{BC}(J_{AC}) = 0$; $J_{AB} > J_{BC}(J_{AC})$; $|\delta_A - \delta_B|$, $|\delta_A - \delta_C|$, $|\delta_B - \delta_C| > J_{AB}, J_{BC}(J_{AC})$. Here $\delta$ are chemical shifts of the signals of the corresponding nuclei, J-the constants of a spin-spin interaction between these nuclei. In the NMR arsenal there is a wide set of impulse radio-frequency sequences which, in principle, can secure (in accordance with the Hadamard's transformation) a selective action on the spins of molecular systems. From the point of view of the perspectives for the use of NMR for solving the problem discussed here the print problems, apparently, are:

1) obtaining (synthesis) the appropriate elementary quantum (elementary processors)-atoms and molecules with the maximum possible number of spin-spin interactions of the tupe

$$A_{ij}I_iI_j \qquad (1)$$

between spins i and j;

2) integrating the given elementary processors into a spatially ordered structure in which the interactions such as (1) between spins of any other nature ensuring the realization of superpositional coherent states can be implemented. In expression (1) $A_{ij}$ or $J_{ij}$, i. e. constants of interaction between spins in diamagnetic molecules or $a_{ij}$ are constants of an ultrafine bond in paramagnetic systems, $J_i$ and $J_j$ are interacting spins. The specificity of constants J is in diamagnetic elementary processors.

1. Variety in magnitude is from $10^{-2}$ Hz up to thousands Hz.
2. The interval of values is characteristic for the given kind of nuclei.
3. The interacting nuclei (spins) can be separated by one, two or several chemical bonds.

The above specificity J allows the creation of algorithms ensuring a sequence of logic (computing) operations. Constants $a$ can also be used for these purposes. By now a great number of compounds synthesized and isolated from natural raw material is known, the NMR spectra of which are well enough studied. Thus, the experimental basis for elementary processor searches is principally formed. Vinyl derivatives with the general structural formula can be among them:

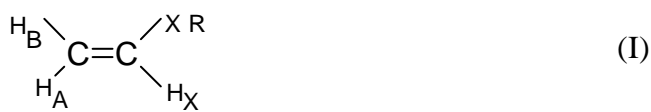

(I)

One of such compounds is given below

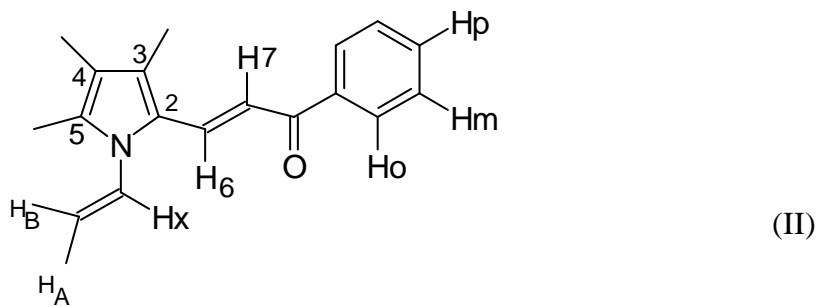

(II)

The following constants of spin-spin coupling (J, Hz) are exhibited in NMR spectra $^1$H and $^{13}$C between the protons ($^1$H-$^1$H):
a) by means of two bonds: A-B=1.22; b) by means of three bonds: X-A=8.53; X-B=15.5; 6-7=15.4; 3-4=3.6; 5-4=2.6; O-M=6.9; M-P=7.0; c) by means of four bonds: 3-5=1.0: O-P=1.8; 6-3=0.6: d) by means of six bonds: 7-5=0.3; A-3=0.35.
between $^1$H and $^{13}$C:
$C_3$-$H_3$=171.2; $C_4$-$H_4$=174.0; $C_5$-$H_5$=185.9; $C_x$-$H_x$=176.5; C-$H_A$=164.3; C-$H_B$=157.8; $C_6$-$H_6$=153.3; $C_7$-$H_7$=157.0; $C_0$-$H_0$=159.6; $C_M$-$H_M$=161.1; $C_P$-$H_P$=161.1.

In fact we have a system of 21 spins which are now characterized by 24 constants. The given example indicates that among the existing compounds one can select molecular objects suitable for the purpose considered here.

Speaking about the prospects of using NMR for quantum calculations it is naturally necessary to take into account the necessary to take into account the specificity of this phenomenon. Firstly, the selection of elementary processors should be done on the basis of the first order spectra, that is the values $\left|J_{ij}\right| << \left|\delta_j - \delta_i\right|$. The realization of the given inequalities infers the application of modern costly spectrometers. Secondly, the NMR phenomenon itself, is speaking generally a slow process that is the result of rather lengthy periods ($T_1$ and $T_2$) of nuclear spins. Therefore, it is logical instead of the re-usable realization of the phenomenon itself to record the deciphered NMR spectrum on some carrier and then to reproduce it with any velocity any time you need (depending on the abilities of the reproducing instrumentation).

The following deviation is pertinent here. The idea of quantum computers put forward by R. Feynman was not the end in itself. Proving it, he meant not individual though important problems (for example, those which can be solved on the basis of the those algorithm) but those which arise in connection with the application of the conceptions of quantum mechanics for describing multicomponent systems. Thus, a problem is a search for the parameters, which characterize the systems, mentioned above. In this case the difficulties connected with their formation were considered to be the result of the impossibility of carrying out the necessary volume of calculations.

The successes achieved by the present time in investigating the structure and dynamics of multielectronic (atomic and molecular) systems have principally changed the situation. In particular, NMR reached such a degree of perfection that on its basis it became possible to solve rather adequately the problems of quantum mechanics (at least, as applied to molecular systems) using a computer. Moreover, the modern state of the experimental data make it possible the solution of which can open new ways of getting information on the peculiarities of a molecule structure.

According to the NMR theory the spin Hamiltonian ($H_S$) describing nuclear interactions in a molecule includes them (linear- $H_S$, bilinear- $H_J$ and square- $H_Q$) as based on the spin functions [4]:

$$H_S = H_Z + H_J + H_Q \quad (2)$$

The linear members in the Hamiltonian are responsible for the perturbation in the system caused by a constant exterior magnetic field and variable radio-frequency field. The result of such interaction is determined, finally, by chemical screening of resonating nuclei in one molecular (multielectronic) system or another the NMR spectrum of which is detected. The component $H_J$ of the Hamiltonian (2) describes indirect (though an electronic medium) and direct interactions of spins among themselves. Finally, $H_Q$ is responsible for the influence of molecular electric fields on the condition of a magnetic resonance. Thus, all the component in (2) contain the information on an electronic medium of resonating nuclei. Exterior conditions are imposed on a nuclear interaction described by $H_Z$ because of chemical screening. The quadrupole interactions with the participation of nuclear spins (in case >1/2) are conditioned by gradient of an electric field which is an averaged characteristic of the given molecule. That is why the motion of the electrons, and consequently, the molecular structure are most adequately reflected in $H_J$. Thus, it is possible to consider that the space in which the spin-spin interaction (SSI) is realized is adequate to its (molecular) dimensions.

It is a well known fact that expressions obtained on the Remzy's formulas and used for calculating constants $J_{ij}$ generally correspond to the following tupe of transition:

$$\Psi(r,E) \rightarrow J \quad (3)$$

Here the left-hand part symbolized the SSI dependence on the spatial and electronic structure of a molecule. In most cases the precise solution of direct problem based on the dependence mentioned above becomes impossible owing to difficulties unique to quantum chemistry. Though for a number of spin-spin interactions it is possible to achieve some quite good compliance between theoretical ($J_T$) and experimental values of constants J [5] using corresponding approaches. The efficiency of theoretical calculations of these constants can at present be considerably increased in case of utilizing the characteristics of atoms which are included into the composition of the investigated molecule and the structure of which is well known. Varying those data one can calculate the values of the SSI constants achieving the necessary compliance between the theory and the experiment with the help of a particular number of iterations. As if was already mentioned above, the modern state of the NMR technique allows the measurement of the most diversified homo- and heteronuclear SSI constants with the highest degree of accuracy. At the same time a great amount of theoretical and experimental material is accumulated concerning the peculiarities of the parameters of atoms (such as C, O, Cl, Br, Si and others) entering the composition different molecules. These two circumstances (in terms of equality $J_{kj}=J_{lk}$) enable the establishment of the spatial structure of molecules. It is convenient to utilize direct constants (C----C, C----N and others) to determine a skeleton and values J by means of two and more bonds - to find out valence characteristics. Here it is pertinent to speak of the method of molecular and iterative building up of molecular systems. As a matter of fact it means the solution of a direct problem:

$$\Psi^*(r,E) \rightarrow J_T^*, \quad (4)$$

where $\Psi^*(r,E)$ are optimized wave functions used to calculate $J_T^*$ close enough to experimental SSI. If calculations are made using a sufficient number of iterations and a big experimental data file one is able, basically, to find out an explicit view of functional dependencies $\Psi(r, E)$, i. e. to transform a direct problem (4) into an inverse one:

$$J \rightarrow \Psi(r,E). \quad (5)$$

The explicit view of wave functions (found in this way) can be used for describing the Hilbert space sized $2^N$. So, there occurs the possibility of operating with a matrix including $4^N$ elements. Each element of the matrix can be compared with the process realizing some algorithm. Hens, a simple processor can become a carrier of data file. If it were possible to connect in a block for example, 100 matrixes in some way, this information would be increased many-fold. Returning to the problem of NMR using for making quantum computers it is possible to make a conclusion that the purpose of this phenomenon is the selection of relevant elementary processors with the ultimate number of interactions such as (1), pin-point values $J_{ij}$ and on this basis – calculation of wave functions and states. One more opportunity is connected with the following circumstance. In accordance with the existing notions the states must "operate" in quantum computers which cannot remain pure (according to the laws of quantum mechanics) since they are subjected to the action (unitary transformations) of macroword on the corresponding degrees of freedom of microparticles (atoms and molecules). In other words, we have to do with the states of microobjects virtual relative to a macroworld. That's why the problem of the way of organizing the process in a macroworld: by detecting some experimental parameter stipulated by quantum (virtual) state or by means of this process modeling with the certain degree of precision becomes not basic. If so, then one can discuss the following variant. The NMR spectra can be applied to the manufacture of the relevant forms (matrices) will be the material for assembling blocks and which are controlled by some methods. In the final, we are dealing with the use of NMR aimed at the expansion of the potentials of modern computers.